\documentstyle[sprocl,epsfig]{article}

\def\NIMA{{\em Nucl. Instrum. Methods} A}

\def\be{\begin{equation}}
\def\ee{\end{equation}}
\def\bea{\begin{eqnarray}}
\def\eea{\end{eqnarray}}
\begin{document}

\def\thefootnote{\fnsymbol{footnote}}
%example \footnote[1]{bla bla bla}

\hfill {\tt LC Note: LC-TOOL-2004-019}

\hfill {\tt arXiv:physics/0409040}

\vspace{33pt}

\title{COMPARISONS OF HADRONIC SHOWER PACKAGES\\}

\author{GEORGIOS MAVROMANOLAKIS~\footnote[1]{~email: {\tt gmavroma@hep.phy.cam.ac.uk} or {\tt gmavroma@mail.cern.ch}} 
and DAVID WARD}

\address{Cavendish Laboratory, University of Cambridge\\
Madingley Road, Cambridge CB3 0HE, U.K.}

\maketitle\abstracts
{
We report on simulation studies comparing various hadronic shower packages. 
Results show that predictions from different models vary significantly, illustrating 
the necessity of testbeam data to resolve the situation. 
}

\vfill
\begin{center}
Proceedings of the International Conference on Linear Colliders \\
LCWS 2004, Paris, 19-23 April 2004
\end{center}
\vfill

\thispagestyle{empty}

\clearpage
%%%..........................................................................

\setcounter{page}{1}

\title{COMPARISONS OF HADRONIC SHOWER PACKAGES\\}

\author{GEORGIOS MAVROMANOLAKIS %~\footnote[1]{email: {\tt gmavroma@hep.phy.cam.ac.uk} or {\tt gmavroma@mail.cern.ch}} 
and DAVID WARD}

\address{Cavendish Laboratory, University of Cambridge, Cambridge CB3 0HE, U.K.}

\maketitle\abstracts
{
We report on simulation studies comparing various hadronic shower packages. 
Results show that predictions from different models vary significantly, illustrating 
the necessity of testbeam data to resolve the situation. 
}

\section{Introduction}

The high precision measurements needed to exploit the physics potential  
of an $e^{+} e^{-}$ Future Linear Collider with 0.5 - 1~TeV center-of-mass 
energy range set strict requirements on performance of vertex, tracking 
and calorimetric detectors. The CALICE Collaboration [\ref{calice}] has 
been formed to conduct the research and development effort needed to 
bring initial conceptual designs for the calorimetry to a final proposal 
suitable for an experiment at the Future Linear Collider. Software 
development and simulation studies play a key role in this effort. Some 
such studies are reported here.

\section{Comparisons of hadronic shower models}

The CALICE Collaboration proposes that both electromagnetic and hadronic  
calorimeters should be highly granular to allow very efficient pattern
recognition for excellent shower separation and particle identification within 
jets and subsequently to provide excellent jet reconstruction efficiency [\ref{calice},\ref{calorimetry}]. 
Prototypes are being constructed and simulation studies are under way to 
support and guide the forthcoming testbeam program. Such studies will help 
to identify regions where testbeams should focus to give answers, resolve 
discrepancies and finally lead to a simulation code with validated and 
reliable predicting power. 

\begin{figure}[t]%------------------------------------------------------------
\vspace{-7pt}
\begin{center}
\begin{tabular}{cc}
\hspace*{-15pt}
\mbox{\epsfig{file=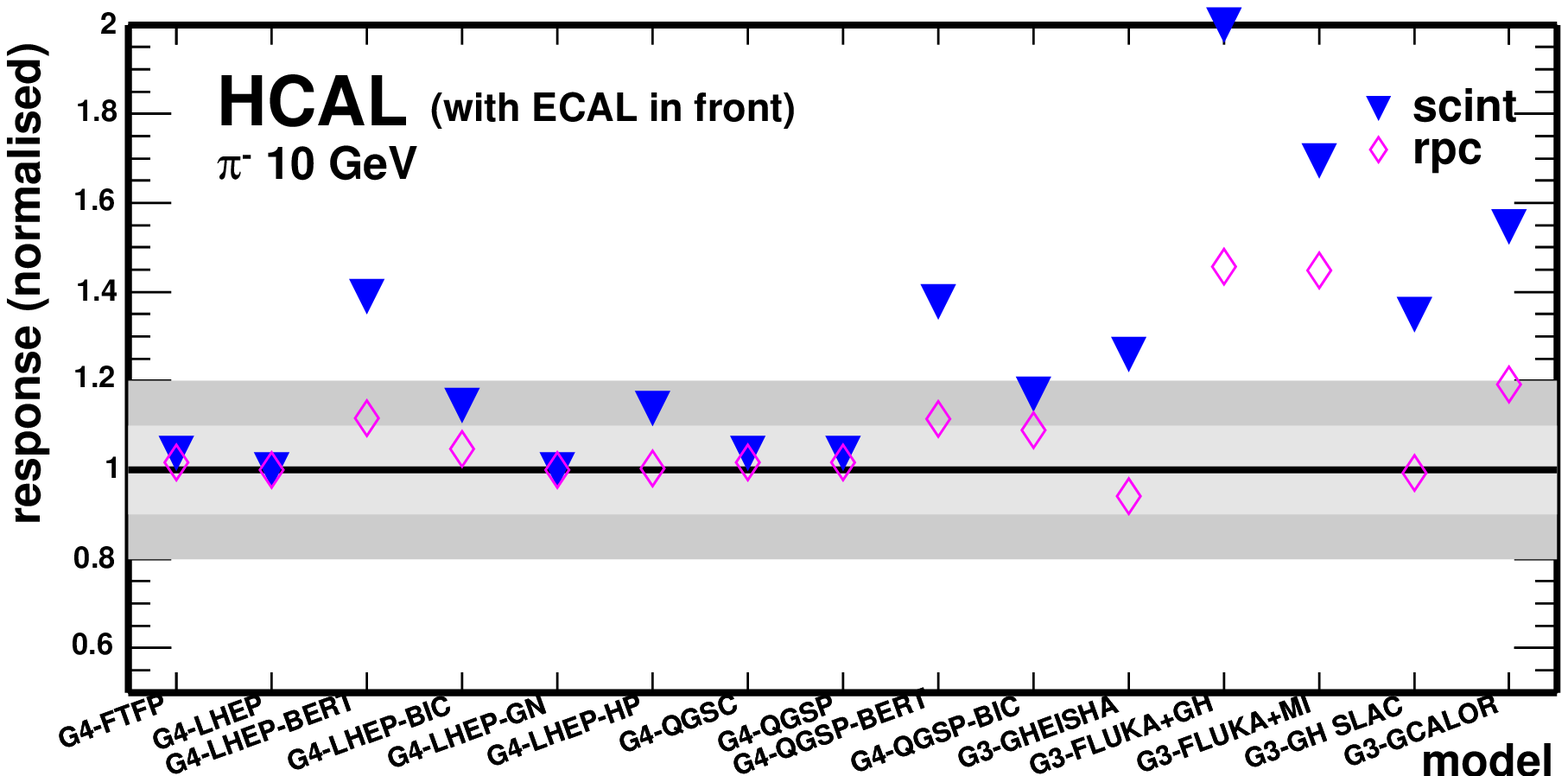,width=6.6cm}} \hspace*{-27pt}&
\mbox{\epsfig{file=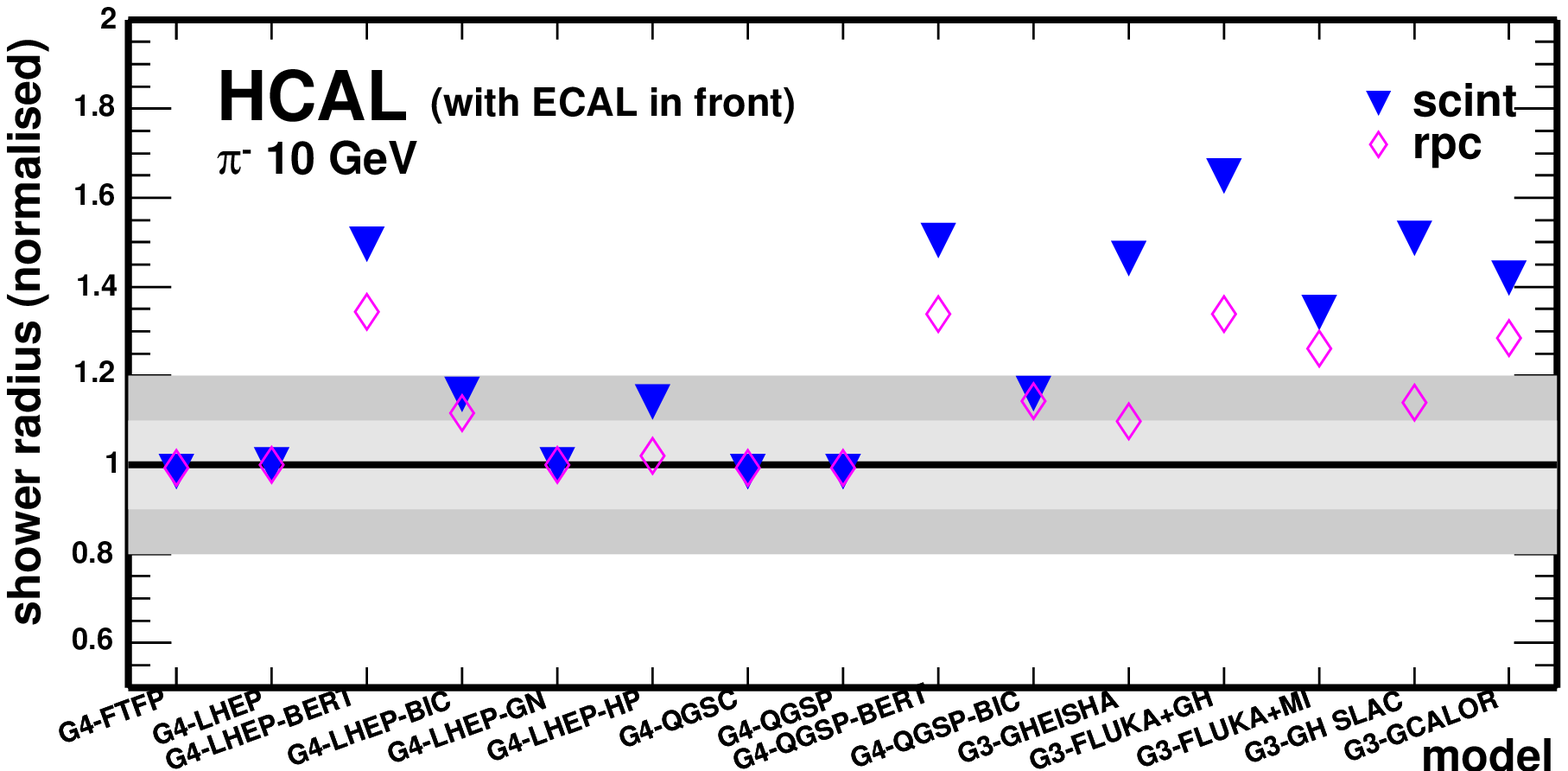,width=6.6cm}} \\
{\small (a)}&{\small (b)}\vspace{-10pt}\\
\end{tabular}
\end{center}
\caption{(a) hadronic calorimeter response, in terms of total number of cells hit, vs hadronic model, 
(b) average shower radius vs model. Results are normalised to the G4-LHEP case, $\pm10\%, \pm20\%$ bands shown to guide the eye.}
\label{fig:one}
%\end{figure}%------------------------------------------------------------------

%\begin{figure}[tb]%------------------------------------------------------------
\begin{center}
\mbox{\epsfig{file=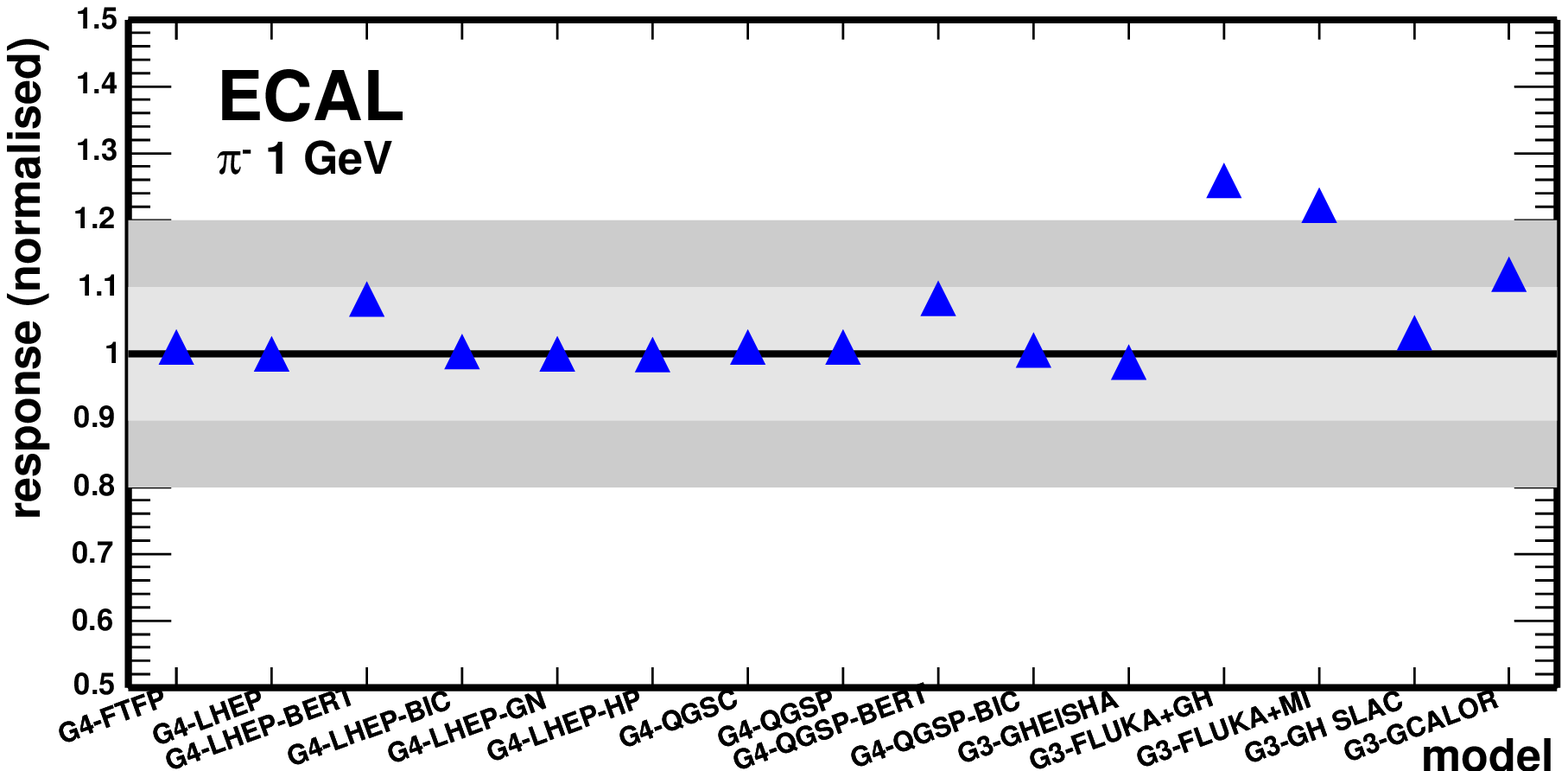,width=6.6cm}}\vspace{-10pt} 
\end{center}
\caption{electromagnetic calorimeter response (total number of cells hit) to incident 
$\pi^-$ at 1 GeV vs hadronic model.}
\label{fig:two}
\end{figure}

In the following we report briefly on systematic comparisons of different 
hadronic shower models. A plethora of models are available within GEANT3~[\ref{geant3}] 
and GEANT4~[\ref{geant4}] simulation frameworks. In 
table~\ref{tab:one} we give a short description of those we have studied. 
In GEANT3 several GHEISHA and FLUKA based models are implemented.
In GEANT4 all models involve GHEISHA; low and high energy 
extensions with intranuclear cascade models and quark-gluon string models 
respectively can be added.
We simulated an electromagnetic calorimeter longitudinally segmented into
30 layers of W of varying thickness as absorber (the first 10 layers at 
1.4~mm thick each, 2.8~mm in the next 10 and 4.2~mm in the final 10) 
interleaved with 0.5~mm Si pads as sensitive material. 
It is read out in 1~cm$^2$ cells.
The hadronic calorimeter consists of 40 layers of Fe absorber, each 18~mm thick,
equipped with scintillator tiles or resistive plate chambers (rpc). 
For the latter version digital readout is envisaged. 
Both versions are simulated as being read out in 1~cm$^2$ cells. 
Detector geometry and material definition were implemented identically in 
both frameworks and their corresponding physics control parameters were 
tuned to produce the same mip peak value for muons. Several experimentally accessible parameters 
predicted by the different models were studied, such as total response, 
response per detector cell, transverse and longitudinal development of showers {\em etc}.
An example, corresponding to incident $\pi^-$ at 10~GeV, is shown in Fig.~\ref{fig:one}. 
Different models predict significantly different HCAL response, Fig.~\ref{fig:one}(a), 
and similarly different shower size, Fig.~\ref{fig:one}(b).
Results for both versions of HCAL are shown.

\begin{table}[t]%-------------------------------------------------------------
\centering
\resizebox{4.7in}{!}{%
\begin{tabular}{ll}
\hline
\hline     
& \vspace{-6pt} \\
{\bf model tag}    & {\bf brief description}\\
& \vspace{-6pt} \\
\hline
& \vspace{-6pt} \\
{\bf G3-GHEISHA}   & GHEISHA, parametrized hadronic shower development \\
                   & \vspace{-6pt} \\
{\bf G3-FLUKA-GH}  & FLUKA, for neutrons with $E<$~20~MeV GHEISHA\\
                   & \vspace{-6pt} \\
{\bf G3-FLUKA-MI}  & FLUKA, for neutrons with $E<$~20~MeV MICAP\\
                   & \vspace{-6pt} \\
{\bf G3-GH SLAC}   & GHEISHA with some bug fixes from SLAC\\
                   & \vspace{-6pt} \\
{\bf G3-GCALOR}    & $E<$~3~GeV Bertini cascade, 3~$<E<$~10~GeV hybrid Bertini/FLUKA, $E>$~10~GeV FLUKA,   \\
                   & for neutrons with $E<$~20~MeV MICAP\\  
                   & \vspace{-6pt} \\
\hline		     
                   & \vspace{-6pt} \\
{\bf G4-LHEP}      & GHEISHA ported from GEANT3\\
                   & \vspace{-6pt} \\
{\bf G4-LHEP-BERT} & $E<$~3~GeV Bertini cascade, $E>$~3~GeV GHEISHA\\
                   & \vspace{-6pt} \\
{\bf G4-LHEP-BIC}  & $E<$~3~GeV Binary cascade, $E>$~3~GeV GHEISHA\\
                   & \vspace{-6pt} \\
{\bf G4-LHEP-GN}   & GHEISHA + gamma nuclear processes\\
                   & \vspace{-6pt} \\
{\bf G4-LHEP-HP}   & as G4-LHEP, for neutrons with $E<$~20~MeV use evaluated cross-section data \\
                   & \vspace{-6pt} \\
{\bf G4-QGSP}      & $E<$~25~GeV GHEISHA, $E>$~25~GeV quark-gluon string model \\
                   & \vspace{-6pt} \\
{\bf G4-QGSP-BERT} & $E<$~3~GeV Bertini cascade, 3~$<E<$~25~GeV GHEISHA, $E>$~25~GeV quark-gluon string model \\
                   & \vspace{-6pt} \\
{\bf G4-QGSP-BIC}  & $E<$~3~GeV Binary cascade, 3~$<E<$~25~GeV GHEISHA, $E>$~25~GeV quark-gluon string model \\
                   & \vspace{-6pt} \\
{\bf G4-FTFP}      & $E<$~25~GeV GHEISHA, $E>$~25~GeV quark-gluon string model with fragmentation ala FRITJOF\\ 
                   & \vspace{-6pt} \\
{\bf G4-QGSC}      & $E<$~25~GeV GHEISHA, $E>$~25~GeV quark-gluon string model \\
& \vspace{-6pt} \\
\hline
\hline
\end{tabular}
}%
\vspace{-0pt} 
\caption{a brief line of description per studied model.}
\label{tab:one}
\end{table}%-------------------------------------------------------------------

In general, our observations from such studies can be summarised by the following:
1) predictions of FLUKA based models are definitely different from those of GHEISHA ones. 
2) The treatment of low energy neutrons is important especially for the scintillator HCAL
and as expected has little effect on a gaseous detector (HCAL rpc).  
3) Intranuclear cascade models also play a crucial role. 
%
%4) scintillator HCAL is more sensitive to model variation than rpc version.
%
4) ECAL standalone with total depth of about 1~$\lambda_{I}$ 
may have some discriminating power with low energy incident hadrons, 
as can be seen in Fig.~\ref{fig:two}. Further detailed studies are under way, 
waiting to be confronted with testbeam data.

%------------------------------------------------------------------

\section{Conclusion}

Simulation studies reveal significant discrepancies among packages, thus  
preventing model independent predictions on calorimeter performance and reliable 
detector design optimization. This underlines the necessity and the importance 
of an extensive testbeam program to resolve the situation and reduce the current
large uncertainty factors.

\end{document}